\documentclass[aps,prb,a4paper,reprint,showkeys,showpacs,superscriptaddress]{revtex4-2}
\usepackage{textcomp}
\usepackage{natbib}
\usepackage{amsmath}
\usepackage{longtable}
\usepackage{rotating}
\usepackage{amssymb}
\usepackage{enumerate}
\usepackage{array}
\usepackage{color}
\usepackage{booktabs,microtype}
\usepackage{graphicx}
\usepackage[margin=1in]{geometry}
\usepackage{hyperref}
\definecolor{dark-red}{rgb}{0.9,0.15,0.15}
\definecolor{dark-blue}{rgb}{0.15,0.15,0.4}
\definecolor{medium-blue}{rgb}{0,0,0.5}
\hypersetup{
	colorlinks=true,
	citecolor=blue,
	linkcolor=red,
	filecolor=blue,      
	urlcolor=blue,
}

\usepackage[charter,greekuppercase=italicized]{mathdesign}

\graphicspath{{./}{figures/}}

\begin{document}

\title{Magnetic order and spin dynamics across the ferromagnetic quantum critical point in Ni\boldmath{$_{1-x}$}Mo\boldmath{$_{x}$}}

\author{H. K. Dara}
\thanks{These authors contributed equally.}
\affiliation{Institute of Physics, Sachivalaya Marg, Bhubaneswar 751005, India}
\affiliation{Homi Bhabha National Institute, Anushakti Nagar, Mumbai 400085, India}

\author{S. S. Islam}
\email{shams.islam@psi.ch}
\thanks{These authors contributed equally.}
\affiliation{PSI Center for Neutron and Muon Sciences CNM, 5232 Villigen PSI, Switzerland}

\author{A. Magar}
\affiliation{School of Physics, Indian Institute of Science Education and Research Thiruvananthapuram-695551, India}

\author{D. Patra}
\affiliation{Department of Physics, Indian Institute of Technology Madras, Chennai 600036, India}

\author{H. Luetkens}
\affiliation{PSI Center for Neutron and Muon Sciences CNM, 5232 Villigen PSI, Switzerland}

\author{T. Shiroka}
\affiliation{PSI Center for Neutron and Muon Sciences CNM, 5232 Villigen PSI, Switzerland}
\affiliation{Laboratorium f\"{u}r Festk\"{o}rperphysik, ETH Z\"{u}rich, CH-8093 Z\"{u}rich, Switzerland}

\author{R. Nath}
\email{rnath@iisertvm.ac.in}
\affiliation{School of Physics, Indian Institute of Science Education and Research Thiruvananthapuram-695551, India}

\author{D. Samal}
\email{dsamal@iopb.res.in}
\affiliation{Institute of Physics, Sachivalaya Marg, Bhubaneswar 751005, India}
\affiliation{Homi Bhabha National Institute, Anushakti Nagar, Mumbai 400085, India}


\begin{abstract}
Realizing a quantum critical point (QCP) in clean ferromagnetic (FM) metals has remained elusive due to the coupling of magnetization to the electronic soft modes that drive the transition to be of first order. However, by introducing a suitable amount of quenched disorder, one can still
establish a QCP in ferromagnets. In this study, we ascertain that the itinerant ferromagnet Ni$_{1-x}$Mo$_{x}$ exhibits a FM QCP at a critical doping of $x_c \simeq 0.125$. Through magnetization and muon-spin relaxation measurements, we demonstrate that the FM ordering temperature is suppressed continuously to zero at $x_c$, while the magnetic volume fraction remains $100\%$ up to $x_c$, indicating a second-order phase transition. The QCP is accompanied by a non-Fermi liquid behavior, as evidenced by the logarithmic divergence of the specific heat and the linear temperature dependence of the low-temperature resistivity. Our findings reveal a minimal effect of disorder on the critical spin dynamics of Ni$_{1-x}$Mo$_{x}$ at $x_c$, highlighting it as one of the rare systems to exhibit a clean FM QCP.
\end{abstract}

\maketitle
Quantum phase transitions (QPTs) are the key to understanding the ground-state
properties and the low-energy emergent collective excitations in a wide class of quantum materials, ranging from high-$T_c$
superconductors and heavy-fermion compounds to low-dimensional systems~\cite{Sachdev29,Sachdev33,Gegenwart186,Sachdev2000}. A QPT is driven by
strong quantum fluctuations between competing ground states
in the vicinity of a quantum critical point (QCP)~\cite{Sachdev29,Sachdev33}.
A QCP can be attained by suppressing a second-order transition at finite
temperature towards $T=0$\,K by varying non-thermal control parameters,
such as chemical doping ($x$), external pressure ($p$), and magnetic
field ($H$), which can lead to novel quantum states, as non-Fermi liquid (NFL) behavior
or unconventional superconductivity~\cite{Gegenwart186,Svanidze011026,Stewart797,Lohneysen532,Lohneysen1015,Schuberth485,Pfleiderer1551}.

Over the years, magnetic QCPs accompanied by NFL behavior have been
extensively studied at the boundary between the antiferromagnetic (AFM)
and paramagnetic (PM) phases in heavy-fermion metals~\cite{Gegenwart186,Stewart797,Lohneysen532}. On the other hand, QCPs in ferromagnetic (FM) metals have proven far more difficult to obtain, as they are inherently unstable~\cite{Brando025006,Belitz4707}. The reason for it is ascribed to the coupling
of the order parameter (magnetization) to the low-energy spin
fluctuations. This coupling drives the phase transition to be of first order (i.e., discontinuous) or triggers the onset of a spin-density-wave (SDW) phase with spatially modulated magnetic order/glassy phase at a finite temperature~\cite{Brando025006,Belitz4707,Kirkpatrick2012universal,Chubukov147003,Conduit207201}. Indeed, clean $3d$-electron based weak itinerant ferromagnets, like ZrZn$_{2}$~\cite{uhlarz2004quantum}, FeGe~\cite{pedrazzini2007metallic}, and MnSi~\cite{pfleiderer1997magnetic,pfleiderer2004partial}, exhibit a first-order transition when the $T_c$ is suppressed to sufficiently low temperature under applied external pressure. Interestingly, the theoretical proposal by Belitz, Kirkpatrick, and Vojta (BKV) on disorder dependence of the FM-QPT suggests that it is possible to restore the second-order FM-QCP in metals by introducing an appropriate level of quenched disorder that suppresses the low-energy fluctuations~\cite{Belitz4707,Sang207201,Kirkpatrick214407}. However, in the strong disorder limit, there is the possibility of observing a Griffiths- or a phase separation that may hinder the realization of a clean FM-QCP~\cite{Vojta2003,Vojta2006rare,Demko2012,Ubaid066402,Wang267202,Uemura29,Westerkamp2009}.

So far there are a few itinerant $d$-electron systems, such as, UCo$_{1-x}$Fe$_{x}$Ge~\cite{Huang237202}, Mn$_{1-x}$Cr$_{x}$Si~\cite{Mishra144436}, (Mn,Fe)Si~\cite{goko2017restoration}, and Ni$_{1-x}$Rh$_{x}$~\cite{Huang117203},
reported to exhibit a clean FM-QCP based on the BKV formalism. In the context of Ni-based binary alloys, efforts have been made to explore QCPs by diluting the Ni magnetic interactions by alloying it with either
paramagnetic or antiferromagnetic metals in the Ni$_{1-x}$M$_{x}$ (M = V, Cr, Rh, Pd, etc.) alloys~\cite{Ubaid066402,Wang267202,Vishvakarma205803,Vishvakarma14,Huang117203,Nicklas4268,Kalvius2014}. 
While Ni$_{1-x}$V$_{x}$~\cite{Ubaid066402,Wang267202} and Ni$_{1-x}$Cr$_{x}$~\cite{Vishvakarma205803,Vishvakarma14} do not feature a QCPs, but rather a quantum Griffiths phase (QGP), there is no definitive proof whether a clean QCP exists in Ni$_{1-x}$Pd$_{x}$, 
as it is susceptible to Ni clustering that leads to inhomogenous alloys~\cite{Kalvius2014}. 
Remarkably, Ni$_{1-x}$Rh$_{x}$ ($x_{c} = 0.375$) is the only member
of the Ni-based binary alloys reported to exhibit a clean
disorder-induced second-order FM-QCP upon dilution of the $d$-electron
magnetic site, based on the BKV formalism~\cite{Huang117203}.
Clearly, the scarcity of $d$-electron-based magnets with clean
second-order FM-QCP drives the search for new candidates capable
of providing new insights into the effects of quenched disorder on the FM-QCP.
Expanding the range of such materials is also essential to identifying
the key factors that enable the formation of a FM-QCP.
Compared to more demanding 
experimental techniques, such as applied external pressure or magnetic field, 
the controlled chemical substitution is a more feasible and versatile
approach to examine the above phenomena.

In this letter, we report the discovery of a clean FM QCP in Ni$_{1-x}$Mo$_{x}$ based on the compelling evidence of (\textit{i}) a continuous suppression of $T_c$ up to the critical concentration ($x_c \simeq 0.125$) via a second-order phase transition, (\textit{ii}) the breakdown of Fermi-liquid behavior across the critical concentration, as evident from the resistivity and heat-capacity data, and (\textit{iii}) the absence of a phase separation close to $x_c$, as revealed by the {\textmu}SR data.

Polycrystalline Ni$_{1-x}$Mo$_{x}$ $(0 \leq x \leq 0.16)$ samples were prepared by arc melting Ni and Mo. Magnetization measurements were carried out using a SQUID magnetometer (MPMS, Quantum Design). Heat capacity and resistivity (current $I = 50$~mA and frequency $f = 2.1$~Hz) were measured using a physical property measurement system (PPMS, Quantum Design). The muon-spin relaxation ({\textmu}SR) experiments under zero field (ZF) and longitudinal field (LF), the latter being applied along the initial muon-spin polarization, were performed at the Dolly spectrometer ($\pi$E1 beamline) at the Paul Scherrer Institute, Villigen, Switzerland. In the current work, 
all the ZF/LF muon-spin polarization curves are plotted after normalizing the respective asymmetries. Details about the experimental methods and sample characterization are provided in the Supplementary Material (SM)~\cite{supple}.

\begin{figure}
	\centering
	\includegraphics[width=\columnwidth]{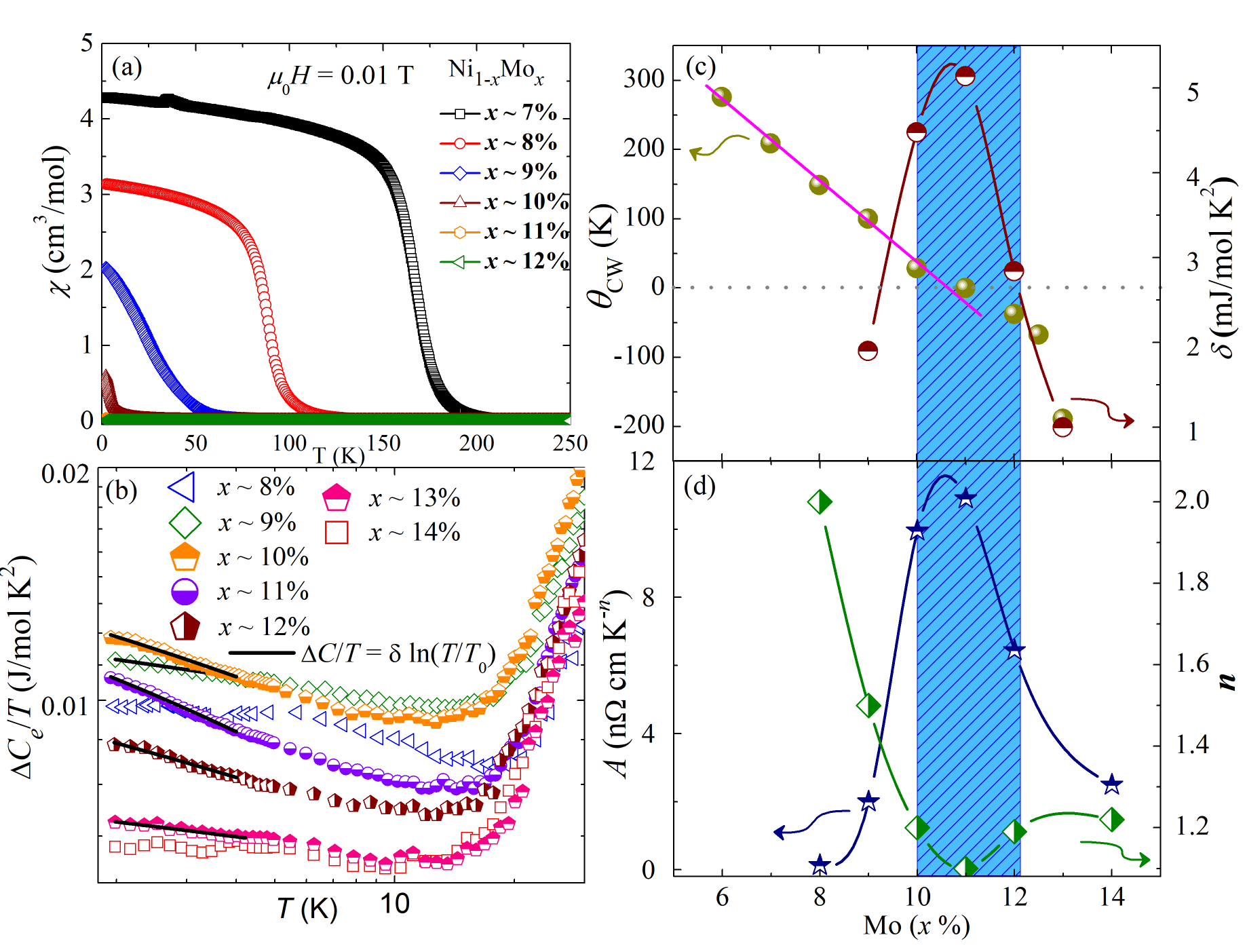}
	\caption{\label{Fig1}(a) DC magnetization $\chi(T)$ measured in a magnetic field of $\mu_{0}H=0.01$\,T. (b) log-log plot of $\Delta C_{e}/T$ vs $T$ for different $x$ values. (c) and (d): the fitting parameters $\theta_{\rm CW}$, $\delta$, $A$, and $n$ are plotted as a function of $x$.
 The hatched area emphasizes the quantum critical region.}
\end{figure}

Ni [bulk Ni has a face-centered cubic (fcc) structure] 
is reported to exhibits a second order PM to FM transition at $T_c \simeq 627$\,K~\cite{Kraftmakher448}. From the temperature dependent magnetic susceptibility $\chi(T)$ we observe that upon substituting Mo
in place of Ni, the FM transition temperature decreases continuously down to 3\,K for $x = 0.10$,
as depicted in Fig.~\ref{Fig1}(a). For $x > 0.11$, $T_c$ is suppressed
down to the lowest accessible temperature of 2\,K [see Fig.~S2(b)~\cite{supple}].  
The Curie-Weiss temperature ($\theta_{\rm CW}$) and the effective paramagnetic moment ($p_{\rm eff}$),  obtained from a modified Curie-Weiss (CW)-law fit to the $\chi^{-1}(T)$ data, exhibit a clear decrease with increasing $x$ (see SM~\cite{supple}), suggesting a reduction in FM correlations as a consequence of the magnetic dilution via the Mo-for-Ni replacement.

The crossover of $\theta_{\rm CW}$ from positive to negative [see Fig.~\ref{Fig1}(c)] implies that the critical concentration ($x_c$) is around 0.12. Our experimental findings are consistent with the theoretical prediction that $T_c$ may approach zero for $x \simeq 0.12$~\cite{Ghosh11773,Singh2478}. 
The itinerant FM character of Ni$_{1-x}$Mo$_{x}$ alloys can be
understood by estimating the Rhodes-Wohlfarth ratio, $RWR = p_{\rm eff}/p_{\rm s}$, 
with $p_{\rm s}$ the spontaneous magnetic moment~\cite{Rhodes247}. For $x = 0.10$ ($T_c\sim 3$~K), $RWR$ is about $\sim 38$ ($p_{\rm eff}$ and $p_{\rm s}$ values are calculated to be 0.87~$\mu_{\rm B}/f.u.$ and 0.023~$\mu_{\rm B}/f.u.$, respectively). This is much larger than unity, indicative
of the weak itinerant FM character of the Ni$_{1-x}$Mo$_{x}$ alloys near $x_c$~\cite{Santiago373002}. In addition, the isothermal magnetization $M(H)$ curves were measured at $T = 3$\,K for different values of $x$ [see Figs.~S2(c),(d)~\cite{supple}]. For $x < 0.11$, 
the magnetization curves exhibit a typical ferromagnetic behavior.
On the other hand, for $x > 0.11$, a pa\-ra\-mag\-net\-ic-like behavior is observed. 

Next, we discuss the departure of the resistivity and heat capacity data from a Fermi-liquid (FL) behavior around $x_c$. The normalized tem\-per\-a\-ture\--de\-pen\-dent electrical resistivity [$\rho(T)/\rho_{\rm 300K}$] measured in zero-field for $x = 0.08$ to 0.14 is depicted in Fig.~S4~\cite{supple}. All the compositions exhibit a typical metallic behavior. The residual resistivity $\rho_{0}$ values are of the order of
{\textmu}$\mathrm{\Omega}$\,cm~\cite{supple}, suggesting minimal disorder
effects on the quantum critical behavior of Ni$_{1-x}$Mo$_{x}$. To examine the electron correlation effect, the data in the low-temperature regime (2 to 75~K) were fitted by $\rho(T)=\rho_{0}+AT^{n}$, where $\rho_{0}$ is the residual resistivity, $A$ is the generalized FL coefficient, and $n$ is a temperature exponent. The variation of the fit parameters $A$ and $n$ with $x$ is shown in Fig.~\ref{Fig1}(d).
As expected for a FL case, the exponent $n$ is close to 2 for $x=0.08$.
As $x$ increases, the $n$ value decreases (to less than 2), to reach a minimum ($n = 1.10$) at $x = 0.11$, and then increases again with $x$. The decrease in $n$ clearly indicates the breakdown
of the FL behavior, here driven by fluctuations in the order parameter.
Concurrently, $A$ also increases with $x$ as it approaches the
$x_c$ value, to exhibit a maximum for $x \simeq 0.11$.

The electronic contribution to heat capacity ($\Delta C_{e}$) is
obtained by subtracting the lattice contribution $C_{\rm lat}$ from
the total heat capacity ($C$) (see SM~\cite{supple}) and is shown
in Fig.~\ref{Fig1}(b) for different $x$ values. 
For $x= 0.08$ and 0.14, both far away from $x_c$, $\Delta C_{e}/T$ remains nearly constant below $\sim 5$~K, an indication of FL behavior. On the other hand, for samples with intermediate concentrations ($x= 0.09$ and 0.13), $\Delta C_{e}/T$ features a logarithmic increment. 
This diverging feature in the low-$T$ regime is a fingerprint of the NFL-type behavior near the critical concentration, here fitted using $\Delta C_{e}/T= - \delta \ln(T/T_{0})$~\cite{Nicklas4268,Vishvakarma14,Sangeetha2019}. The obtained $\delta$ values, shown in Fig.~\ref{Fig1}(c), portray a well-defined maximum around the critical concentration $x \simeq 0.11$ that correlates with the parameters obtained from the resistivity data analysis.

\begin{figure*}
	\centering
	\includegraphics[width=\linewidth]{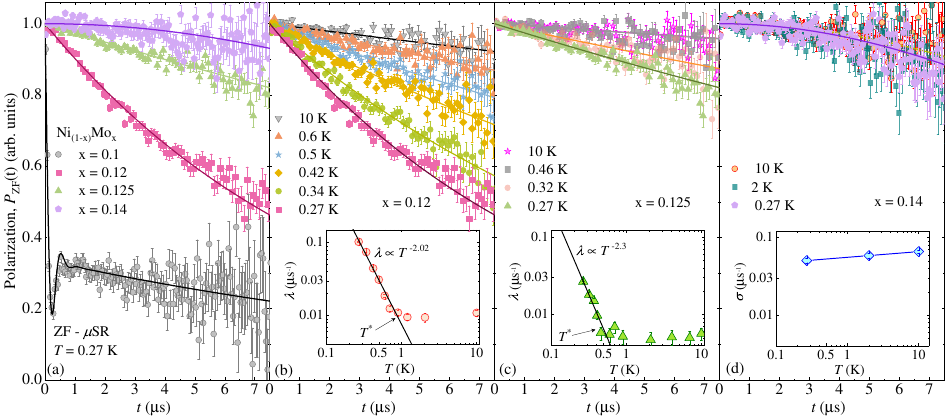}
	\caption{\label{Fig2} (a) ZF muon spin polarization curves, $P_{\rm ZF}(t)$ for samples with different Mo concentrations, $x = 0.1$--0.14, measured at $T = 0.27$\,K. Temperature evolution of the $P_{\rm ZF}(t)$ curves for (b) $x=0.12$, (c) $x = 0.125$, and (d) $x = 0.14$. Solid lines represent the fits as described in the text. Temperature dependence of the muon-spin relaxation rates for  $x = 0.12$ (inset of b), $x = 0.125$ (inset of c), and for $x = 0.14$ (inset of d). Solid-line fits indicate a power-law behavior.}
\end{figure*}

In order to gain microscopic insight into the regime near the FM-QCP, muon-spin relaxation ({\textmu}SR) experiments were carried out down to 270\,mK on samples across $x_c$
(i.e., $x = 0.10$, 0.12, 0.125, and 0.14). First, by analyzing the
$T$-dependent weak transverse field (wTF) and zero-field (ZF) {\textmu}SR data, 
we reveal 
the nature of the magnetic ground state for $x = 0.10$ (see SM for the details~\cite{supple}). Our results identify 
a homogeneous static magnetic order without any phase separation for $x = 0.10$.
Next, we follow the evolution of the static- and dynamic
electronic properties with doping. 
Figure~\ref{Fig2}(a) depicts the gradual change in the ZF muon-spin
polarization, $P_{\rm ZF}(t)$, 
in Ni$_{1-x}$Mo$_{x}$ at 0.27\,K for different Mo compositions: $x = 0.1$, 0.12, 0.125, and 0.14. The early-time fast-relaxing component and the spontaneous oscillation associated with the static magnetic order observed for $x=0.1$, vanish completely and an exponential type relaxation behavior takes over for $x\geq0.12$ near the FM QCP. For $x = 0.14$, $P_{\rm ZF}(t)$ changes it nature from exponential to a pronounced Gaussian type, as expected for a nonmagnetic ground state, where muon spin-depolarization occurs only through interactions with the nuclear dipole moments. For a comparison, in Fig.~\ref{Fig2}(b-d), we show the $T$-dependent $P_{\rm ZF}(t)$ curves, measured down to 0.27\,K, for $x=0.12$, 0.125, and 0.14, respectively. A systematic decrease in the $T$-dependent relaxation is observed with increasing Mo content, from $x = 0.12$ to 0.14. The observation of a distinct single-component exponential relaxation for $x = 0.12$ and 0.125 provides strong evidence for a spatially uniform magnetic state throughout the sample volume. It also reinforces the conjecture that, for samples with $x \geq 0.12$, the magnetically ordered volume fraction $V_{\rm ord}$ drops to zero without any sign of phase separation~\cite{Uemura29, Frandsen12519, Huang117203}. The $P_{\rm ZF}(t)$ curves for the $x = 0.12$ and 0.125 samples are modeled by a simple exponential function: $P_{\rm ZF}(t) = e^{-\lambda t}$,  with the extracted muon-spin relaxation rates $\lambda (T)$ being plotted in the insets of Fig.~\ref{Fig2}(b) and (c), respectively. We find that, for $x = 0.12$ (0.125), $\lambda$ stays nearly $T$-independent down to $T^{*}\simeq1$~K~(0.5~K), below which it increases smoothly following a power law $\lambda\propto T^{-\alpha}$ with $\alpha \simeq 2.02~(2.3)$ due to the slowing down of spin fluctuations. 
The reason behind the observed anomalous spin dynamics (with $\alpha >1$) is currently not fully understood.  
However, it is worth mentioning that such a power-law behavior of $\lambda(T)$, with exponents $\alpha \geq 1$, is seen also in other quantum critical systems, such as YFe$_2$Al$_{10}$ ($\alpha \sim$1.2)~\cite{Huang155110}, Nb$_{1.0117}$Fe$_{1.9883}$ ($\alpha \sim 1.16$)~\cite{Willwater134408}, Nb$_{1.0055}$Fe$_{1.9945}$ ($\alpha \sim 2.7$)~\cite{Willwater134408}, Ce\-Pd$_{0.84}$\-Ni$_{0.16}$Al ($\alpha \sim 1.7$)~\cite{Ishant023112}, etc.
Finally, for $x=0.14$, the $P_{\rm ZF}(t)$ curves could be nicely
fitted with a static Gaussian Kubo-Toyabe relaxation,
$G_{\rm KT} =  \frac {1}{3} + \frac {2}{3} \left(1-\sigma^2 t^2\right) e^{-\frac {1}{2}\sigma^2 t^2}$. The Gaussian relaxation rate $\sigma$ exhibits a nearly $T$-independent behavior, as shown in the inset of Fig.~\ref{Fig2}(d).

\begin{figure*}
	\centering
	\includegraphics[width=\linewidth]{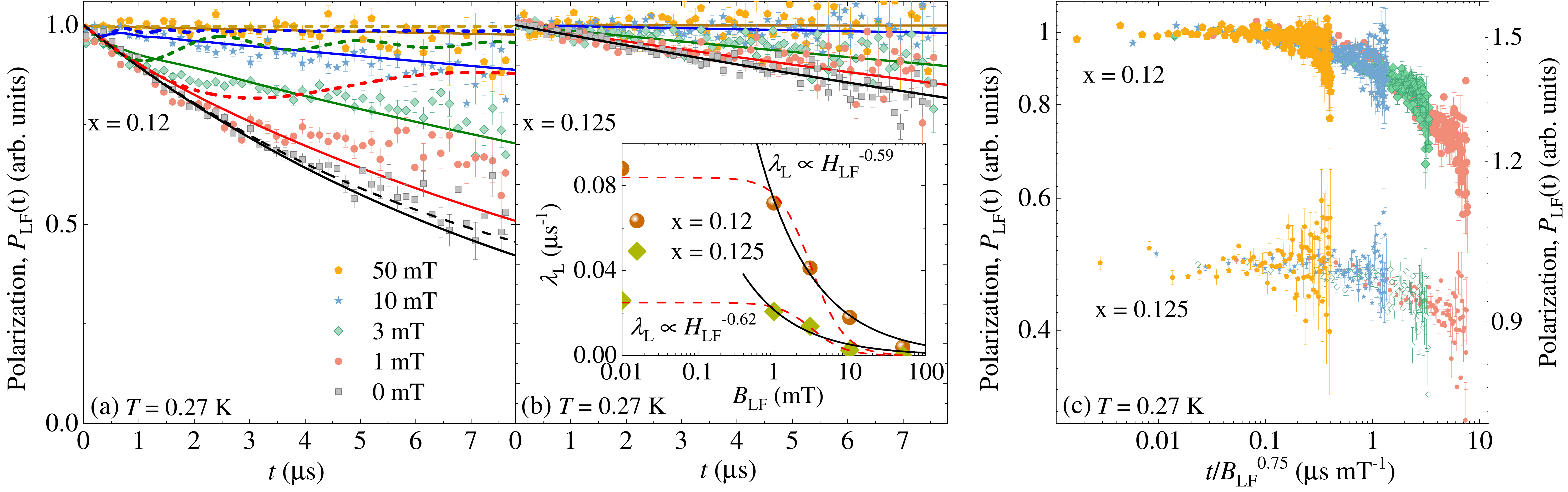}
	\caption{\label{Fig3} $T$-dependence of LF muon-spin polarization curves, $P_{\rm LF}(t)$s for (a) $x=0.12$, and (b) $x=0.125$. In (a), the dashed and solid lines represent simulated depolarization curves using static and dynamic Lorentzian Kubo-Toyabe functions, respectively. Inset of (b) longitudinal field dependence of the dynamic muon-spin relaxation rate $\lambda_{\rm L}$. The dashed lines are Redfield fits, whereas, the solid lines are power-law fits. (c) Muon-spin polarization curves measured under various $B_{\rm LF}$ as function of the scaling variable $t/B_{\rm LF}^{0.75}$ for $x = 0.12, 0.125$.}
\end{figure*}

To distinguish between the static- and dynamic magnetic correlations in Ni$_{1-x}$Mo$_{x}$ close to the FM QCP, we performed a series of {\textmu}SR experiments at different longitudinal fields (LFs) at 0.27\,K for $x=0.12$ and 0.125. The measured LF depolarization curves $P_{\rm LF}(t)$ are shown in Fig.~\ref{Fig3}(a) and (b), respectively. Clearly, the muon spins are completely decoupled in an applied field $B_{\rm LF} \simeq 50$\,mT. To better understand the spin dynamics for $x = 0.12$, the ZF data was fitted to the theoretical Kubo–Toyabe functions for both static (dashed black line) and dynamic (solid black line) relaxation~\cite{Kubo1967}. Thereafter, using the fitted parameters, the static (dashed lines) and dynamic (solid lines) relaxations are simulated for various longitudinal fields. If the muon-spin depolarization were caused only by a static distribution of local fields, the experimental $P_{\rm LF}(t)$ curves for higher longitudinal fields would resemble the simulated static (dashed lines) relaxation curves~\cite{Spehling140406, Amato2024}, which is clearly not the case. Instead, we observe 
only a partial decoupling. Whereas, the predictions for a dynamical relaxation (solid lines) show a much better agreement (at least at long times).
For $x=0.125$, all the $P_{\rm LF}(t)$ curves 
follow a simple exponential behavior. To quantify the evolution of the
spin fluctuation rate with the applied LF, the $P_{\rm LF}(t)$ curves
of both $x=0.125$ and $x=0.12$ (shown in Fig.S7 of SM~\cite{supple}) were
fitted by an exponential function, $P_{\rm LF}(t) = e^{-\lambda_{\rm L} t}$.
The field-dependence of the obtained muon-spin relaxation rate,
$\lambda_{\rm L}$, for both $x=0.12$ and 0.125, is
shown in the inset of Fig.~\ref{Fig3}(b). For applied longitudinal fields $B_{\rm LF} > 1$\,mT, $\lambda_{\rm L}$ decreases gradually towards zero following a power law $\lambda_{\rm L}\propto B_{\rm LF}^{-\beta}$, with an exponent $\beta \simeq 0.59~(0.62)$ for $x = 0.12$ (0.125).

We recall that, the LF-dependence of the muon-spin depolarization is a direct probe of the Fourier transform of the spin-spin
autocorrelation function $q(t)=\langle s_{i}(t)\cdot s_{i}(0)\rangle$. According to theory, depending on the behavior of $q(t)$, one can
deduce the nature of a magnetic system. For instance, for a magnetically homogeneous system, $q(t)$ is a simple exponential.
In case of an inhomogeneous system, $q(t)$ exhibits a power-law or a stretched exponential behavior. To quantify the autocorrelation time $\tau_c$, the LF-dependence of $\lambda_{\rm L}$,
for both $x = 0.12$ and 0.125, is approximated by the Redfield formalism [shown as red solid line in the inset of Fig.~\ref{Fig3}(b)]~\cite{Spehling140406, Sarkar121111, Adroja014412}:
\begin{equation}
	\begin{aligned}
		\label{Redfield}
		\lambda_{\rm L} (B_{\rm LF}) = 2\Delta^2 \tau_c / [1 + 2 \gamma_{\mu}^2 B_{\rm LF}^2 \tau_c^2].
	\end{aligned}
\end{equation}
Here, $\Delta^2 = \gamma_{\mu}^2 \langle B_{\rm loc}^2\rangle$ describes the time-averaged second moment of the local fluctuating magnetic field distribution $H_{\rm loc}$ at the muon stopping site with a single fluctuation time $\tau_c$.  
When $k_{\rm B}T\gg\hbar\omega_{\mu}$, the fluctuation-dissipation theorem correlates $\tau_c$ with the $q$-independent dynamic susceptibility, $\chi^{\prime\prime}(\omega_{\mu})$, through the relation $\tau_c = \left(k_{\rm B}T/\mu_{\rm B}^2\right)[\chi^{\prime\prime}(\omega_{\mu})/\omega_{\mu}]$. A satisfactory fit to the $\lambda_{\rm L} (B_{\rm LF})$ data for $x = 0.12$ yields $\Delta^2 \simeq 0.11$~MHz and $\tau_c = 3.8\times10^{-7}$~s, becoming $\Delta^2 \simeq 0.03$\,MHz and $\tau_c = 3.7\times10^{-7}$\,s
for $x = 0.125$. The resulting $\tau_c$ values exceed those of
spin-glass systems (e.g., $10^{-8}-10^{-11}$\,s, in AuFe or CuMn~\cite{Uemura546}). Yet, the $\tau_c$ values of Ni$_{1-x}$Mo$_{x}$ are similar to those reported for the few other FM QCP systems, such as YbNi$_4$(P$_{1-x}$As$_x$)$_2$~\cite{Spehling140406, Sarkar121111}, CePd$_{0.15}$Rh$_{0.85}$~\cite{Adroja014412}, etc. In general, such $\tau_c$ values indicate very slow critical spin fluctuations.

Another method to determine the behavior of $q(t)$ is the time-field scaling analysis of the $P_{\rm LF} (t)$ curves~\cite{Ishida184401, Lausberg216402, MacLaughlin066402}. For both the Ni$_{1-x}$Mo$_{x}$ samples with $x = 0.12$ and 0.125, the $P_{\rm LF} (t)$ curves are found to obey the time-field scaling relation $P_{\rm LF} (t, B_{\rm LF}) = P_{\rm LF} (t/B_{\rm LF}^{\gamma})$ with $\gamma = 0.75$, as shown in Fig.~\ref{Fig3}(c) and (f). For such low applied LFs, we expect the field dependence to be due to a change in $\omega_{\mu}$ rather than to the effects of LF on $q(t)$. From the $\gamma$ value, one can distinguish different types of $q(t)$ behavior. For instance,
when $q(t)$ follows a power law, $q(t)= ct^{-\alpha}$, one expects $\gamma=1-\alpha<$1, whereas when $q(t)$ is a stretched exponential,  $q(t) = c~e^{-\left(\lambda t\right)^\beta}$, $\gamma=1+\beta>$1. In our case, the observed scaling exponent $\gamma = 0.75$ indicates that the spin-spin autocorrelation
function can be well approximated by a power law rather than by a stretched
exponential. Our time-field scaling analysis of the LF data strongly suggests
that, for both samples, cooperative- and critical spin fluctuations
exist, rather than a distribution of local spin-fluctuation rates~\cite{MacLaughlinS4479,Ishida184401,Lausberg216402}. 

In order to substantiate our observation of quantum critical fluctuations and to understand the role of disorder behind the NFL behavior observed in Ni$_{1-x}$Mo$_{x}$ with $x = 0.12$ and 0.125, we compare our {\textmu}SR data with those 
of the few other materials showing NFL behavior near quantum criticality. Prominent examples in this category are CeFePO~\cite{Lausberg216402}, CePd$_{0.15}$Rh$_{0.85}$~\cite{Adroja014412}, UCu$_{5-x}$Pd$_x$, $(x = 1, 1.5)$~\cite{MacLaughlin066402, MacLaughlinS4479}, etc. There is no 
universal consensus, whether in these systems the
critical fluctuations arise as a consequence of disorder or they are
intrinsic to the QCP. In this respect, our ZF and LF-$\mu$SR spectra
in Ni$_{1-x}$Mo$_{x}$ with $x = 0.125$ mostly show a single-exponential type behavior, similar to that observed in YbNi$_{4}$(P$_{1-x}$As$_x$)$_2$~\cite{Spehling140406, Sarkar121111}, YFe$_2$Al$_{10}$~\cite{Huang155110}, and Nb$_{1.0117}$Fe$_{1.9883}$~\cite{Willwater134408}. This may indicate that the degree of disorder is not as stringent 
as in other critical NFL materials~\cite{Lausberg216402, Adroja014412, MacLaughlin066402} hence, suggesting that disorder-driven mechanisms, such as the Griffiths phase~\cite{Wang267202}, cannot be the 
driving mechanism behind the observed NFL behavior in our case. Therefore, Ni$_{1-x}$Mo$_{x}$ for $x=0.125$ can be categorized as a comparatively cleaner quantum critical system than its counterparts~\cite{Wang267202}. Note that, very recently, ac susceptibility measurements have shown a spin-glass-like transition in Ni$_{1-x}$Mo$_{x}$ with $x = 0.095$~\cite{Hsu416524}, which could be due to a phase segregation (formation of clusters) 
of Mo during the sample preparation. In contrast, our {\textmu}SR data suggest a critical spin dynamics without any indication of spin-glass order near the FM QCP in the $x = 0.12$ and 0.125 samples.

In summary, based on thermodynamic- and {\textmu}SR measurements, we have presented convincing experimental evidence on the occurrence of a second-order FM-QCP in Ni$_{1-x}$Mo$_{x}$ alloys at a critical concentration $x_c = 0.125$. The ZF- and LF-{\textmu}SR data exclude either 
a long- or short-range order, as well as any significant influence of disorder on the slow critical fluctuations at the FM QCP. Furthermore, the low-temperature resistivity- and heat-capacity data suggest the breakdown of Fermi-liquid behavior near $x_c$, due to the presence of quantum critical spin fluctuations. Our findings further emphasize the importance of maintaning a low amount of disorder in order to achieve a clean QCP in the itinerant ferromagnet-based alloys.

H. K.\ Dara thanks G. Markandeyulu and V. Srinivas for granting access to the sample preparation facilities at IIT Madras and fruitful discussions.
R.N.\ and A.M.\ acknowledge financial support from SERB, India, bearing sanction
Grant No.~CRG/2022/000997.

\bibliography{NiMo}

\end{document}


\title{Supplementary Material to\\
Magnetic order and spin dynamics across the ferromagnetic quantum critical point in Ni\boldmath{$_{1-x}$}Mo\boldmath{$_{x}$}}

\author{H. K. Dara}
\thanks{These authors contributed equally.}
\affiliation{Institute of Physics, Sachivalaya Marg, Bhubaneswar 751005, India}
\affiliation{Homi Bhabha National Institute, Anushakti Nagar, Mumbai 400085, India}

\author{S. S. Islam}
\email{shams.islam@psi.ch}
\thanks{These authors contributed equally.}
\affiliation{PSI Center for Neutron and Muon Sciences CNM, 5232 Villigen PSI, Switzerland}

\author{A. Magar}
\affiliation{School of Physics, Indian Institute of Science Education and Research Thiruvananthapuram-695551, India}

\author{D. Patra}
\affiliation{Department of Physics, Indian Institute of Technology Madras, Chennai 600036, India}

\author{H. Luetkens}
\affiliation{PSI Center for Neutron and Muon Sciences CNM, 5232 Villigen PSI, Switzerland}

\author{T. Shiroka}
\affiliation{PSI Center for Neutron and Muon Sciences CNM, 5232 Villigen PSI, Switzerland}
\affiliation{Laboratorium f\"{u}r Festk\"{o}rperphysik, ETH Z\"{u}rich, CH-8093 Z\"{u}rich, Switzerland}

\author{R. Nath}
\email{rnath@iisertvm.ac.in}
\affiliation{School of Physics, Indian Institute of Science Education and Research Thiruvananthapuram-695551, India}

\author{D. Samal}
\email{dsamal@iopb.res.in}
\affiliation{Institute of Physics, Sachivalaya Marg, Bhubaneswar 751005, India}
\affiliation{Homi Bhabha National Institute, Anushakti Nagar, Mumbai 400085, India}

\maketitle

\section{Sample Preparation and X-ray Diffraction}
\begin{figure*}[htb!]
	\centering
	\includegraphics[width=0.7\textwidth]{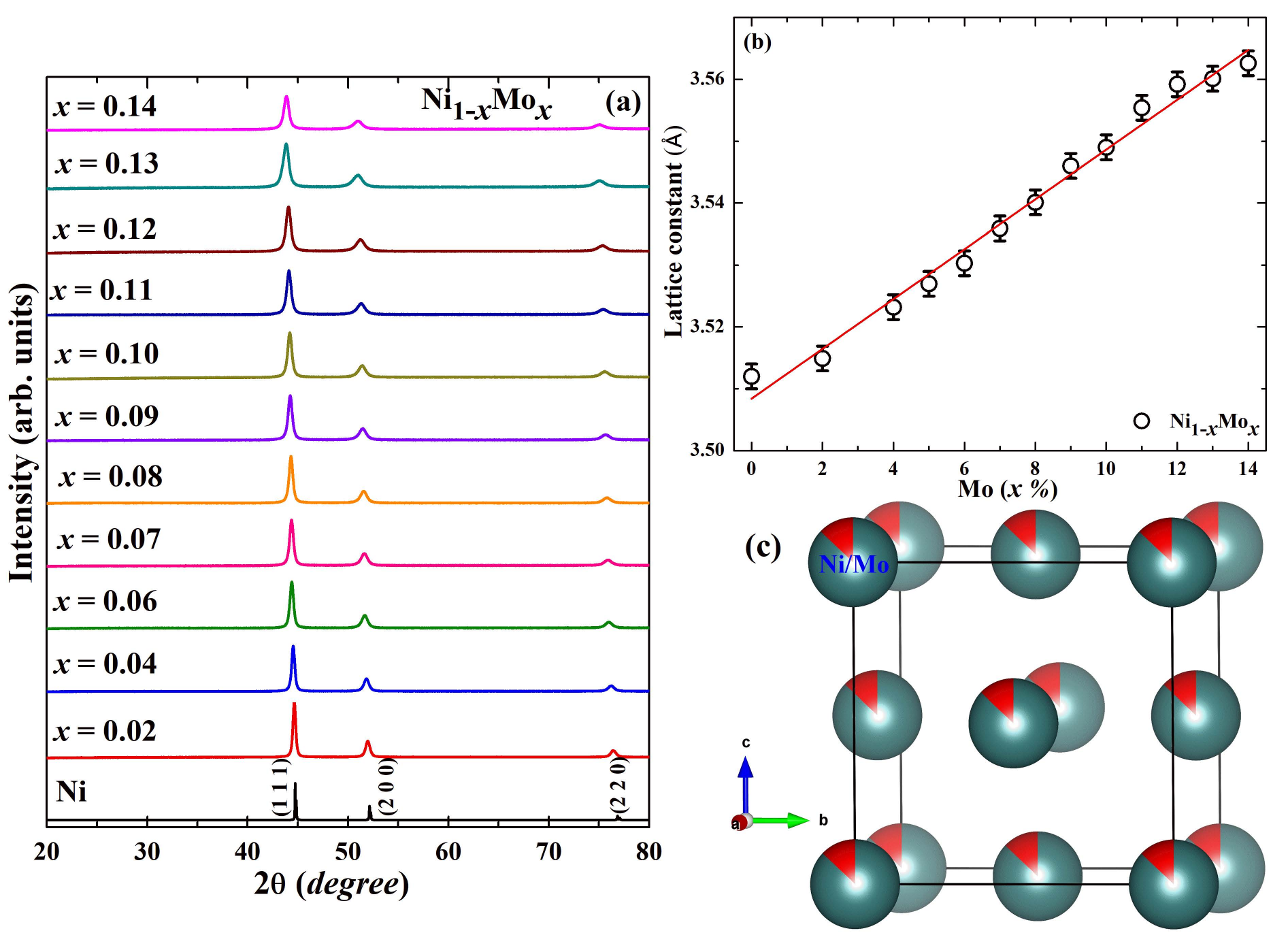}
	\caption{\label{FigS1}(a) Room temperature x-ray diffraction patterns
	for different concentrations of Mo. (b) Variation of the
	lattice constant with Mo content $x$. The solid red line represents a linear fit to the data. (c) Unit cell of Ni$_{1-x}$Mo$_{x}$, with the colored spherical sectors representing the partial substitution of Mo.}
\end{figure*}

Polycrystalline samples of Ni$_{1-x}$Mo$_{x}$ ($x = 0.0$--0.14) were prepared
by arc-melting the constituent elements Ni and Mo, with a weight loss below 0.2$\%$. The ingots were re-melted several times by flipping the sample after
each melting to ensure its chemical homogeneity, followed by annealing at 1270~K.
To confirm the
crystal structure and phase purity of the sample, powder x-ray diffraction patterns were recorded at room temperature using a PANalytical x-ray diffractometer (XRD) with Cu K${\alpha}$ ($\lambda = 1.5406$\,\AA) radiation [see Fig.~\ref{FigS1}(a)]. The Rietveld refinement of the XRD patterns was
carried out using the FullProf software \cite{Rodriguez55}.
All the XRD patterns could be refined well using the
cubic space group $Fm\overline{3}m$ and the lattice constant  
$a = 3.512$\,\AA\ of metallic Ni as a starting parameter.
As shown in Fig.~\ref{FigS1}(b), the lattice constant increases linearly
with the Mo concentration $x$ and is consistent with the Vegard's law \cite{Vegard17}. This suggests that Mo is indeed substituted at the Ni site. The increase in lattice constant with
$x$ can be ascribed to the larger atomic radius of Mo (1.363~\AA)
compared to Ni (1.246~\AA).

\section{DC magnetization}
\begin{figure*}
	\centering
	\includegraphics[width=0.7\textwidth]{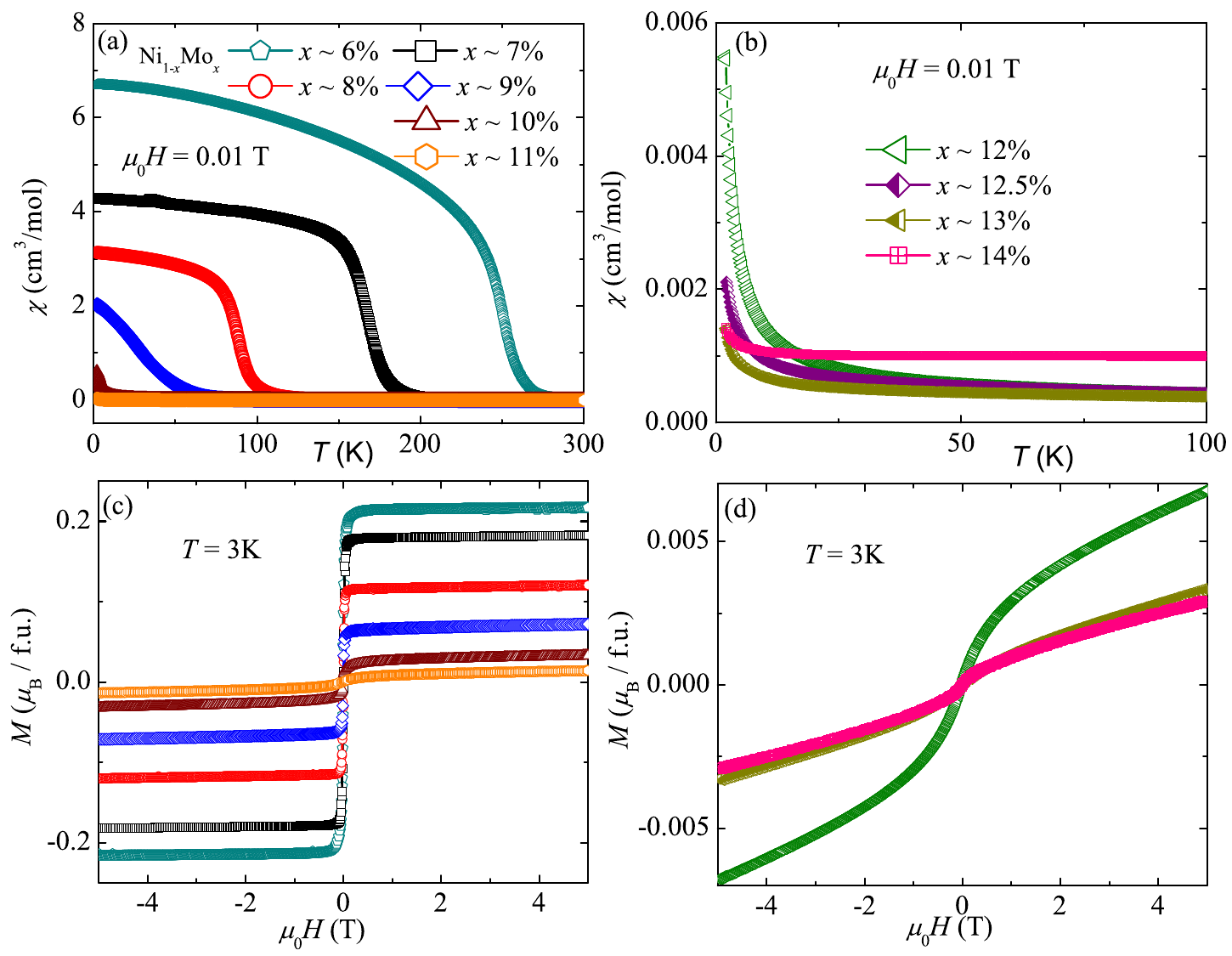}
	\caption{\label{FigS2}(a) and (b): DC magnetic susceptibility vs $T$ for different Mo concentrations ($x = 6$\% to 14\%) measured at $\mu_0 H = 0.01$~T. (c) and (d): Isothermal magnetization curves measured at $T = 3$~K for the same compositions.}
\end{figure*}
Figures~\ref{FigS2}(a) and (b) present the temperature dependent
dc magnetic susceptibility $\chi(T)$ ($\equiv M/H$, where $M$ is the magnetization and $H$ is the applied field) for all the Ni$_{1-x}$Mo$_{x}$
samples ($x = 0.06$--0.14), measured at $\mu_0 H = 0.01$~T. As $x$ increases, the ferromagnetic transition ($T_\mathrm{c}$) shifts systematically towards lower temperatures and is suppressed below 2\,K for $x>11$\%.
Therefore, for higher concentrations $x$, in Fig.~\ref{FigS2}(b)
we show $\chi(T)$ only in the low-temperature regime to highlight
the purely paramagnetic nature of the compounds.
Similarly, $M$-vs-$H$ magnetic isotherms, measured at $T=3$\,K for
all the compositions ($x = 0.06$--0.14) are shown in Figs.~\ref{FigS2}(c)
and (d). For $x$ values below 11\%, 
$M$ features a sudden jump with increasing field, to saturate completely
at low applied fields, as expected for a ferromagnetic material.
However, for $x \ge 12$\%, $M$ does not show a complete saturation, rather, it behaves more like a paramagnet. This is a further indication that, for $x \ge 12$\%,
$T_\mathrm{c}$ is suppressed below 3~K. For all the FM samples ($x< 11$\%), the $M$ data in high fields were fitted by a straight line and extrapolated to zero-field. From the intercept at the $y$-axis, the saturation magnetization ($M_{\rm s}$) was estimated for different values of $x$ and is tabulated in Table~\ref{tab:MCW}.

\begin{figure*}
	\includegraphics[width=1.0\textwidth]{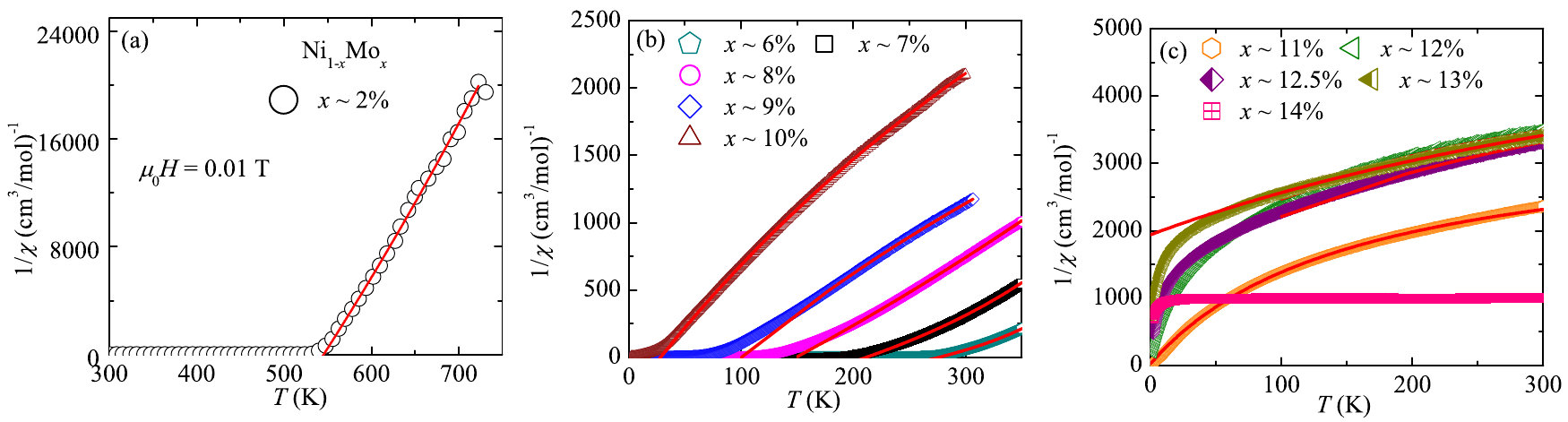}
	\caption{Inverse susceptibility $1/\chi$ vs $T$ for different $x$ values.
	Solid red lines represents fits using a modified Curie-Weiss law.}
	\label{FigS3}
\end{figure*}

The inverse magnetic susceptibility $1/\chi$ as a function of temperature
is plotted in Fig.~\ref{FigS3}. For lower concentrations, at high temperatures,
$1/\chi$ exhibits a linear behavior and is fitted well by the modified
Curie-Weiss law, $\chi = \chi_0 + \frac{C}{T-\theta_{\rm CW}}$, where $C$
is the Curie-Weiss constant, $\theta_{\rm CW}$ represents the characteristic Curie-Weiss temperature, and $\chi_0$ is the temperature independent susceptibility that takes into account the intrinsic core diamagnetism and Pauli paramagnetism. From the value of $C$, the effective moment is calculated as $p_{\rm eff} = \sqrt{3k_{\rm B}C / N_{\rm A}}$, where $N_{\rm A}$ is the Avogadro number and $k_{\rm B}$ is the Boltzmann constant.
The obtained parameters ($\theta_{\rm CW}$, $\chi_0$, and $p_{\rm eff}$) and $p_{\rm s}$ values are summarized in Table~\ref{tab:MCW} for all the compositions. 

\begin{table}
	\centering{}
	\caption{Curie-Weiss temperature $\theta_\mathrm{CW}$,   
	temperature independent susceptibility $\chi_{0}$, spontaneous moment
	$p_\mathrm{s}$, and effective paramagnetic moment $p_\mathrm{eff}$
	for the Ni$_{1-x}$Mo$_{x}$ samples.}
	\label{tab:MCW}
	\setlength\tabcolsep{0.6em}
	\begin{tabular}{*{5}{S[table-format=2.2,table-number-alignment = center]}}  
		\toprule
		{$x$ (\%)} & {$\theta_\mathrm{CW}$ (K)} & {$\chi_{0}$ (cm$^3$/mol)} &
		{$p_{s}$ ($\mu_\mathrm{B}$/f.u.)} & {$p_\mathrm{eff}$ ($\mu_\mathrm{B}$/f.u.)} \\
		\midrule
		0.0  & {--} & {--} &  0.600  & {--} \\ 
		2 & 545.5 &  -5.4$\times$10$^{-6}$ & 0.501 & {--} \\ 
		4 & 466.9 &   2.66$\times$10$^{-6}$ & 0.376 & {--} \\ 
		6 & 276.06 & -17.5$\times$10$^{-4}$ & 0.214 & 2.08 \\ 
		7 & 208.9 & -6.65$\times$10$^{-4}$ & 0.168 & 1.69 \\ 
		8 & 148.9 & -1.11$\times$10$^{-4}$ & 0.115 & 1.39 \\ 
		9 & 100.2 & 1.45$\times$10$^{-4}$ & 0.065 & 1.08 \\ 
		10 & 28.5 & 1.21$\times$10$^{-4}$ & 0.023 & 0.91 \\ 
		11 & 0.14 & 2.82$\times$10$^{-4}$ & {--} & 0.61 \\ 
		12 & -37.42 & 2.05$\times$10$^{-4}$ & {--} & 0.57 \\ 
		12.5 & -67.19 & 1.74$\times$10$^{-4}$ & {--} & 0.50 \\ 
		13 & -189 & 1.52$\times$10$^{-4}$ & {--} & 0.46 \\ 
		\bottomrule
	\end{tabular}
\end{table}

To gain further insight, the isothermal magnetization $M(H)$ curves were
measured at $T = 3$~K for different values of $x$ [see Figs.~S2(c), (d)]. For $x$ values below 0.11,
$M$ saturates quickly at low fields and the magnetization
curves are reminiscent of typical ferromagnets. However, as $x$ increases,
the absolute value of saturation magnetization $M_{\rm s}$ decreases systematically, following a similar trend to that of $T_c$.
At the same time, for $x > 0.11$, the shape of the isotherms is
different, reflecting the paramagnetic behavior of the compounds.


\section{Resistivity}
\begin{figure}
	\includegraphics[width=\columnwidth]{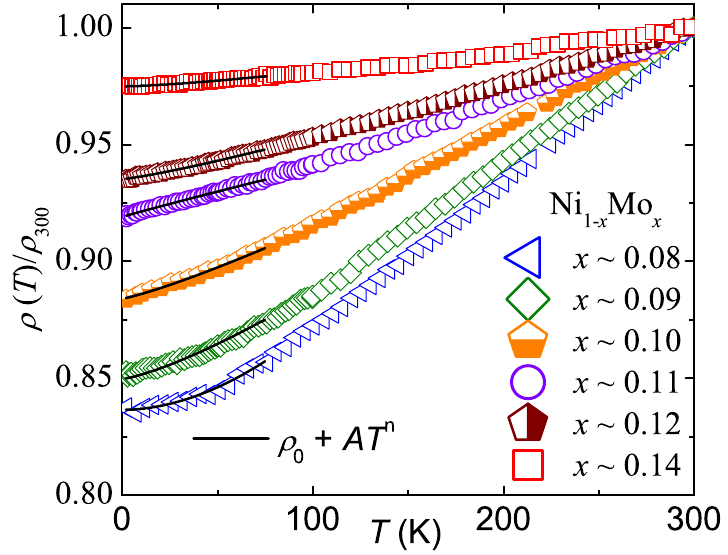}
	\caption{$\rho(T)/\rho_{\rm 300K}$ vs $T$, for $x = 0.08$ to 0.14. The solid line is a fit using $\rho(T)=\rho_{0}+AT^{n}$ in the low temperature regime.}
	\label{FigS4}
\end{figure}

The experimental resistivity data were fitted by a power law,
$\rho=\rho_{0}+AT^{n}$, in the temperature range 2 to 75\,K, as shown
in the Fig.~\ref{FigS4}. The obtained fitting parameters ($\rho_{0}$, \textit{A}, and \textit{n}) are summarized in Table~\ref{tab:resis}. 

\begin{table}
	\centering
	\caption{Residual resistivity $\rho_{0}$, $A$ coefficient,
	and $n$ exponent for Mo concentration $x = 0.08$ to 0.14.}
	\label{tab:resis}
	\setlength\tabcolsep{0.6em}
	\begin{tabular}{*{4}{S[table-format=2.2,table-number-alignment = center]}} 
		\toprule
		{$x$ (\%)} & {$\rho_{0}$ ({\textmu}$\mathrm{\Omega}$\,cm)} &
		{$A$ (n$\mathrm{\Omega}$\,cm$\cdot$K$^{-n}$)}  & {$n$}  \\ 
		\midrule
		8 & 44 & 0.123 & 2 \\ 
		9 & 43 & 3 & 1.5 \\ 
		10 & 60 & 9.96 & 1.2 \\ 
		11 & 72 & 10.9 & 1.10 \\ 
		12 & 82 & 6.44 & 1.19 \\ 
		14 & 109 & 2.49 & 1.22 \\ 
		\bottomrule
	\end{tabular}
\end{table}

\section{Heat Capacity}
\begin{figure}
	\includegraphics[width=\columnwidth]{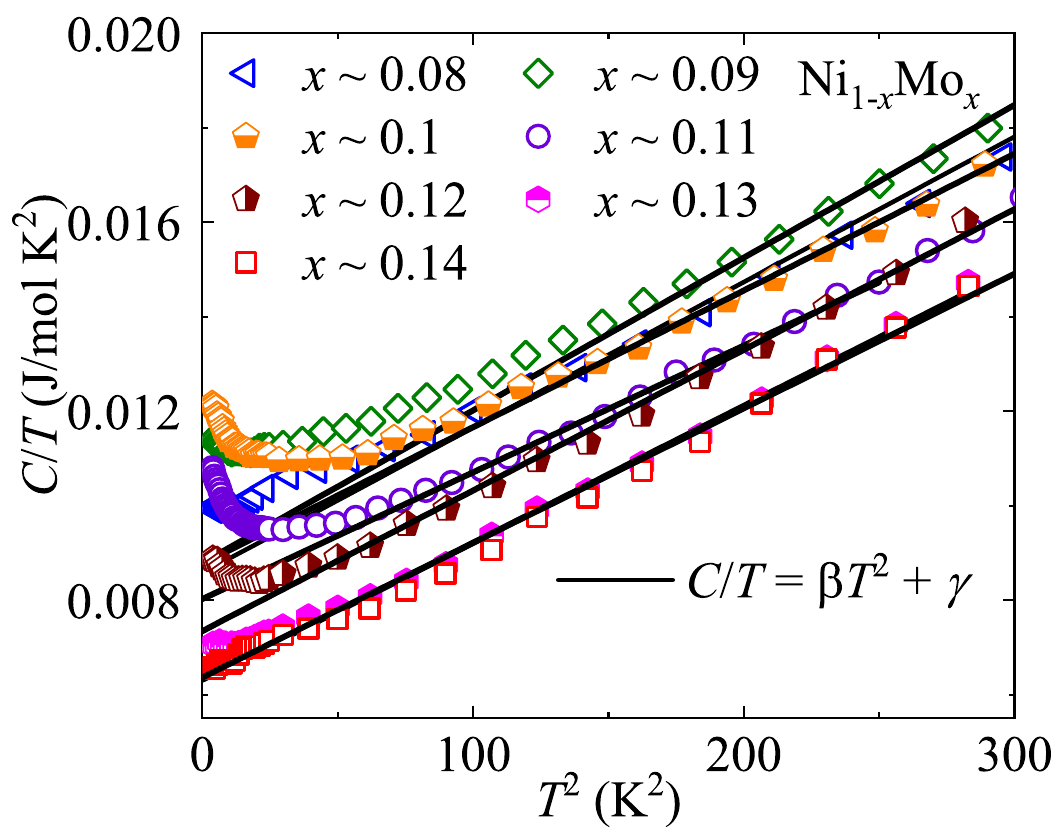}
	\caption{$C/T$ vs $T^2$ plots with the solid lines representing linear fits as described in the text.}
	\label{FigS5}
\end{figure}
The low-$T$ heat capacity ($C$) measured in zero-field for different values of $x$ is presented in Fig.~\ref{FigS5}. Typically, for a metallic system, $C$ has two major contributions
\begin{equation}
	C/T=\gamma +\beta T^{2},
	\label{eq:Hc}
\end{equation}
where the first term accounts for the electronic contribution ($C_{\rm el}$) and the second term is due to the lattice part ($C_{\rm lat}$). In Fig.~\ref{FigS4}, we have plotted $C/T$ vs $T^2$, which shows a linear region for $T>10$~K for all the $x$ values. The obtained parameters ($\gamma$ and $\beta$) from the linear fit in this regime are tabulated in Table.~\ref{tab:HC}. For $T< 10$~K, the data deviate from the fit, a deviation which becomes more pronounced as $x$ approaches $x_c$, likely due to the enhancement of critical fluctuations. The electronic contribution ($\Delta C_{e}$), obtained by subtracting $C_{\rm lat}$ from the total $C$, is shown in Fig.~1(b)
for different $x$ values.
From the fit values obtained for $\beta$ and $\gamma$, one
can evaluate the Debye temperature $\theta_\mathrm{D}$ and the
density of states at the Fermi level [$D(E_\mathrm{F})$] as
\begin{equation}
	\theta_\mathrm{D}=\left (\frac{12\pi^{4}Rn}{5\beta}\right)^{1/3},
	\label{eq:Debye}
\end{equation}
\begin{equation}
	D(E_\mathrm{F})=3\gamma/\pi ^{2}k_\mathrm{B}^{2}N_\mathrm{A}.
	\label{eq:DOS}
\end{equation}
Here, \textit{n} is the number of atoms per formula unit (\textit{n} = 1),
$\gamma$ is the Sommerfield coefficient, $k_{B}$ is
the Boltzmann constant, $N_\mathrm{A}$ is the Avogadro constant,
and $R$ is the gas constant. The estimated parameters $\beta$ and $\theta_{D}$
are tabulated in Table~\ref{tab:HC} along with the $\gamma$ and $\delta$-values.

\begin{table*}
	\centering
	\caption{Variation of $\gamma$, $\beta$, $\delta$, $\theta_{D}$, and
	$D(E_{\mathrm{F}})$ with $x$ for Ni$_{1-x}$Mo$_{x}$ ($x = 0.08$ to 0.14).}
	\label{tab:HC}
	\setlength\tabcolsep{0.6em}
	\begin{tabular}{*{6}{S[table-format=2.2,table-number-alignment = center]}} 
	\toprule
		{$x$ (\%)} & {$\gamma$ (mJ/mol$\cdot$K$^{2}$)} & {$\beta$ (mJ/mol$\cdot$K$^{4}$)} &
		{$\delta$ (mJ/mol$\cdot$K$^{2}$)} & {$\theta_\mathrm{D}$ (K)} &
		{$D(E_\mathrm{F}$) (states/eV$\cdot$f.u.)} \\
	\midrule
		8 & 8.57 & 0.0307 & {--} &  36.97 & 3.67 \\ 
	    9 & 8.7  & 0.0274 & 1.9  & 374.92 & 3.69 \\ 
	    10 & 8.77 & 0.0247 & 4.48 & 388.11 & 3.72 \\ 
	    11 & 8.09 & 0.0282 & 5.14 & 371.34 & 3.43 \\ 
	    12 & 7.34 & 0.0297 & 2.84 & 364.98 & 3.11 \\ 
	    13 & 6.31 & 0.0290 & 1 & 367.89 & 2.68 \\ 
	    14 & 6.36 & 0.0284 & {--} & 370.46 & 2.70 \\
	\bottomrule
	\end{tabular}
\end{table*}

\section{$\mu$SR}
\begin{figure*}
	\centering
	\includegraphics[width=\linewidth]{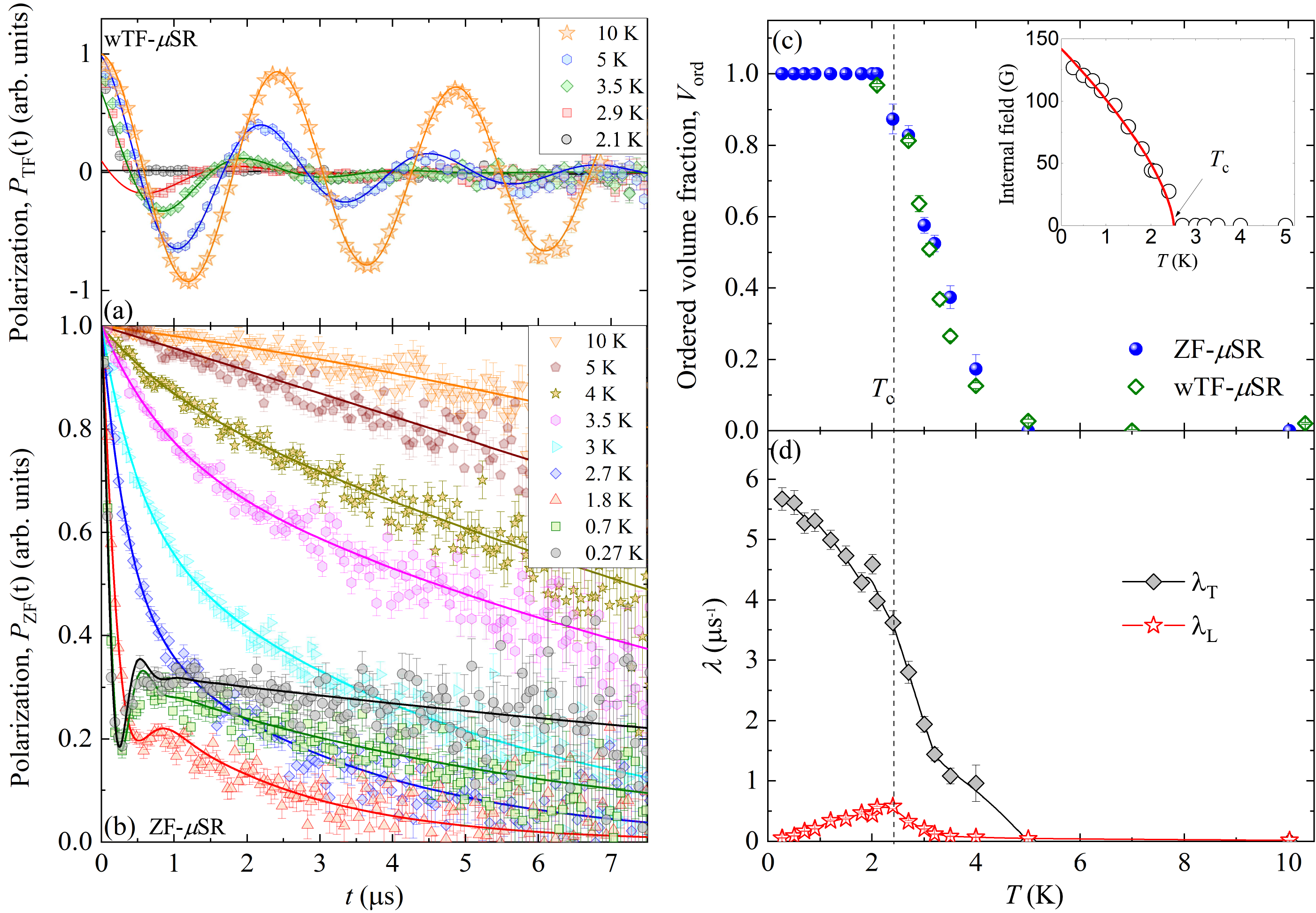}
	\caption{\label{FigS6}{\textmu}SR results in Ni$_{1-x}$Mo$_{x}$,
		for $x = 0.1$. (a) wTF-{\textmu}SR spectra recorded at various temperatures and the corresponding fits (solid lines), performed using Eq.~(\ref{eq:fit_function_1_1}). (b) ZF-{\textmu}SR spectra measured at different temperatures down to 0.27\,K. The solid lines are fits using Eq.~\eqref{fit_function_1}, as described in the text. (c) The temperature dependence of the magnetically ordered volume fraction. The data extracted from weak-TF {\textmu}SR measurements are shown with green diamonds, while those from ZF-{\textmu}SR are shown with blue circles. Inset: evolution of
		the internal magnetic field with decreasing temperature. The red curve represents a mean-field power-law fit. (d) Temperature dependence of the transverse- and longitudinal relaxation rates.}
\end{figure*}
Implanted muons can detect internal fields as low as $10^{-5}$~T, thus being among the most sensitive probes of the absence of frozen disordered magnetic moments in a material~\cite{Amato2024}. Moreover, {\textmu}SR can probe the local order parameter and the magnetically ordered volume fraction independently. We recall that the order parameter is indicative of  the type of thermal phase transition (first/second-order)~\cite{Frandsen12519}. The volume fraction, instead, allows one to unambiguously distinguish between a phase separation and a homogeneous magnetic order~\cite{Uemura29}. Therefore, {\textmu}SR serves as an indispensable local experimental technique to study QCP.

In a typical muon-spin rotation ($\mu$SR) experiment, spin-polarized muons
are implanted into a sample. Implanted muons generally stop at interstitial sites inside the sample, where their spins start to precess about the local magnetic field. The precessing muons decay with an average life time of 2.2\,$\mu$s, producing two neutrinos and a positron. The direction of positron emission lies preferentially along the initial direction of
the muon spin. Thus, the time evolution
of the decay positrons provides direct information about the
muon-spin polarization. Emitted positrons are detected in forward (F) and backward (B) detectors placed along the initial muon-spin polarization direction. Histograms $N_\mathrm{F}(t)$ and $N_\mathrm{B}(t)$ record the number of positrons detected in the two detectors as a function of time following the muon implantation. The time evolution of the muon polarization can be inferred from the normalized difference between $N_\mathrm{F}(t)$ and $N_\mathrm{B}(t)$ and is given by the asymmetry function: 
\begin{equation}
	\label{muon_asymm}
	A(t) = \frac{N_\mathrm{F}(t) - \alpha N_\mathrm{B}(t)}{N_\mathrm{F}(t) + \alpha N_\mathrm{B}(t)} = A_0 P(t).
\end{equation}
Here, $\alpha$ is an experimental calibration constant,
usually determined by applying a weak transverse field (3\,mT, in our case). The quantities $A(t)$ and $P(t)$ depend sensitively on the static spatial distribution and the fluctuations of the magnetic environment of the muons. 

A selection of muon-spin polarization $P_{\rm TF}(t)$ curves,
measured under a weak transverse field (wTF) of 3\,mT with decreasing
temperature (from 10\,K to 2.1\,K) are shown in Fig.~\ref{FigS6}(a).
In a typical wTF-{\textmu}SR experiment, the spins of muons implanted
in a nonmagnetic sample 
precess under a small applied field, producing a long-lived
oscillating signal with maximum amplitude. Conversely, in case of
a magnetic order, the implanted muons experience a broad
field distribution (due to the vector sum of the applied transverse- and the
internal field), resulting in a rapid dephasing or in a
reduction of the oscillating amplitude. The $P_{\rm TF}(t)$
curves for all the temperatures 
were analyzed by using the function:
\begin{equation}
	\begin{aligned}
		\label{eq:fit_function_1_1}
		P_{\rm TF}(t) = P_{\rm TF}(0) e^{-\lambda t} \cos{\left( \omega t +\phi\right)}.
	\end{aligned}
\end{equation}
Here, $P_{\rm TF}(0)$ is the initial polarization (amplitude),
proportional to the paramagnetic volume fraction, $\lambda$ is the
exponential damping rate due to the paramagnetic spin fluctuations,
$\omega$ is the Larmor precession frequency (proportional to the
applied field), and $\phi$ is a phase offset. At 10\,K, $P_{\rm TF}(0)\simeq1$, indicating the lack of a magnetic order. At 2.1\,K, $P_{\rm TF}(0)$
is almost zero, indicating the development of magnetic order in
most of the sample volume. The magnetically ordered volume fraction
$V_{\rm ord}$ is evaluated from the simple relation $V_{\rm ord}=1-P_{\rm TF}(0)$ 
and shown as empty green diamonds in Fig.~\ref{FigS6}(c). The $T$-dependence
of $V_{\rm ord}$ shows a relatively broad transition, from the paramagnetic
state to the magnetic order, with the magnetic- and paramagnetic
regions coexisting in the 2.1--5\,K interval.

Fig.~\ref{FigS6}(b) depicts a few representative 
zero-field polarization curves $P_{\rm ZF}(t)$, measured in
Ni$_{1-x}$Mo$_{x}$ for $x = 0.1$ at different temperatures
(from 10\,K to 0.27\,K). At high temperatures (10\,K), $P_{\rm ZF}(t)$
shows a weakly-decaying Gaussian-like behavior, typically observed in the paramagnetic regime. Here, the dominant depolarization channel is
through the dense array of static nuclear dipole moments rather
than via electronic moments which are fluctuating very rapidly and only mildly depolarizing the implanted muons. Upon cooling below $T \simeq 5$\,K,
$P_{\rm ZF}(t)$ gradually changes its shape, from weakly Gaussian
to exponential-like, indicating the growing contribution
of the electronic moments. As the temperature decreases further,
an additional fast-relaxing component develops at early times
($t \leq 1$\,{\textmu}s). Upon further lowering the temperature,
at $T \leq 2.4$\,K, $P_{\rm ZF}(t)$ eventually develops a fast damped
oscillation, involving $2/3$ of the total asymmetry and
followed by a $1/3$ slow relaxing tail. The reason behind the
$2/3$ oscillating and $1/3$ non-oscillating components is the spatial
averaging of the internal magnetic fields in polycrystalline samples.
Indeed, here $2/3$ ($1/3$) of the internal field components are
oriented perpendicular (parallel) to the direction of the initial muon
spins, causing a precession (no precession) of the muon spin~\cite{Amato2024}.
The observation of the $2/3$- and $1/3$ signal fractions strongly
suggests that 100$\%$ of the sample volume experiences a
homogeneous quasi-static magnetic order without any phase separation~\cite{Spehling140406, Uemura29}. The coherent oscillation with a single well-defined
frequency indicates the presence of a unique dominant muon-stopping site.

\begin{figure*}
	\centering
	\includegraphics[width=0.5\linewidth]{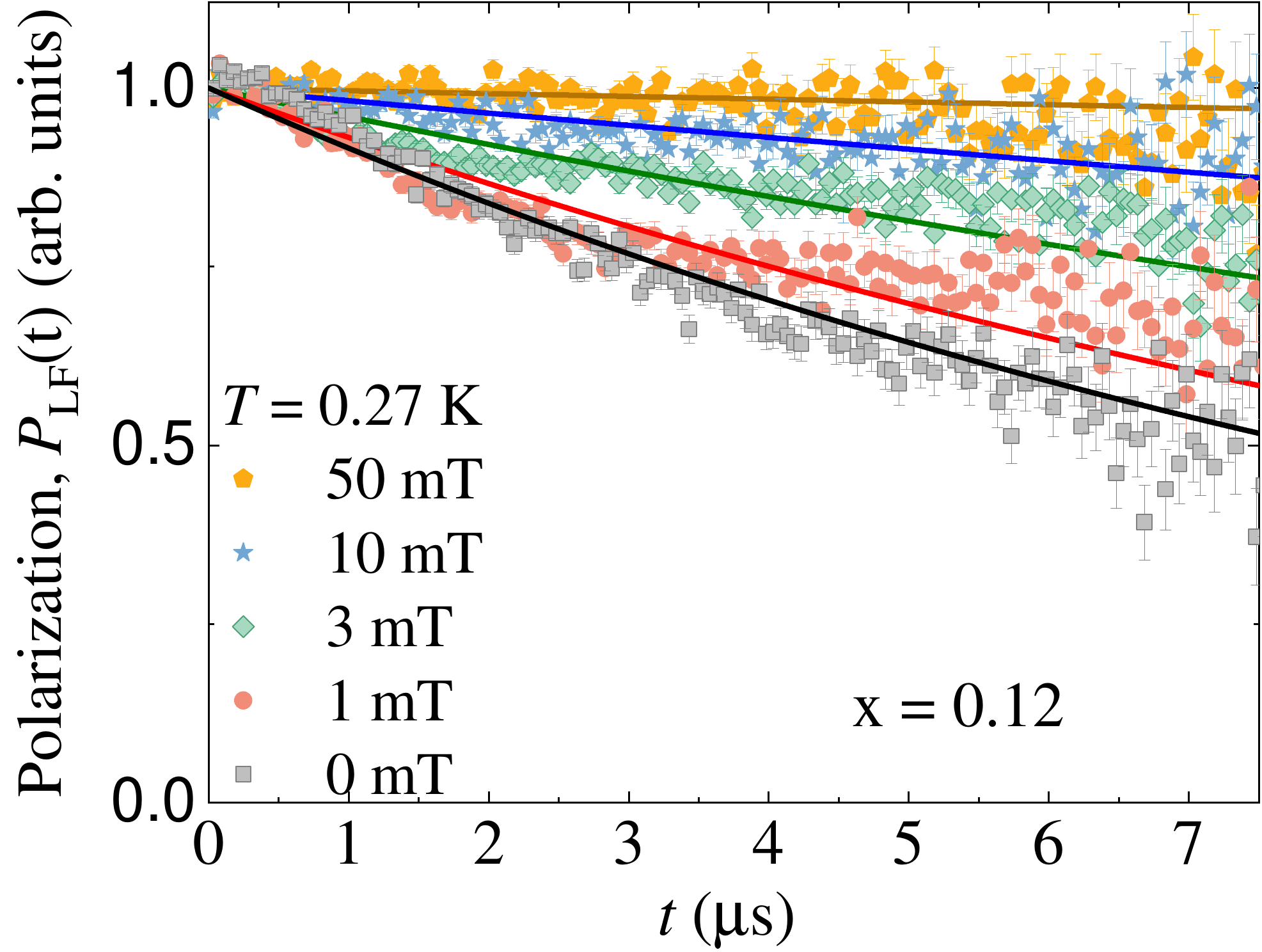}
	\caption{\label{FigS7} LF muon-spin polarization, $P_{\rm LF} (t)$, for x = 0.12 obtained at $T = 0.27$~K, in various applied longitudinal fields. The solid lines are the fits using a simple exponential function.}
\end{figure*}

To describe the ZF-polarization over the whole measured $T$-range,
we used the following fitting model:
\begin{equation}
	\begin{aligned}
		\label{fit_function_1}
		P(t) = V_{\rm ord}~\Big[\frac {2}{3}~{\rm cos}(\gamma_{\mu} B_{\rm int} t)~e^{-\lambda_{\rm T} t} + \frac {1}{3}~e^{-\lambda_{\rm L} t}\Big] \\ + \left( 1 - V_{\rm ord} \right) G_{\rm KT} e^{-\lambda_{\rm L} t}.
	\end{aligned}
\end{equation}
Here, $V_{\rm ord}$ is the magnetically ordered volume fraction,
$B_{\rm int}$ is the local magnetic field at the muon\--stop\-ping site,
while $\lambda_{\rm T}$ and $\lambda_{\rm L}$ are the transverse- and
longitudinal muon-spin relaxation rates. Finally, $G_{\rm KT}$ is
the well-known static Gaussian Kubo-Toyabe depolarization function, 
$ G_{\rm KT} =  \frac {1}{3} + \frac {2}{3} \left(1-\sigma^2 t^2\right)
e^{-\frac {1}{2}\sigma^2 t^2}$. A satisfactory fit to the high-temperature
($T = 10$\,K) data gives $\sigma \simeq 0.04$\,{\textmu}s$^{-1}$, which was then kept fixed for the entire $T$-range while fitting. The magnetically ordered volume
fraction $V_{\rm ord}$, shown in Fig.~\ref{FigS6}(c), increases
gradually with decreasing temperature and reaches 100\% below $T\simeq2.4$~K. Simultaneously, a continuous growth of the order parameter, i.e., the
internal magnetic field $B_{\rm int}$ is observed with decreasing
temperature, as shown in the inset of Fig.~\ref{FigS6}(c), suggesting a
second-order type behavior of the thermal phase transition~\cite{Frandsen12519}.
Although the oscillating signal disappears above $T \simeq 2.4$\,K,
a contribution from the non-oscillating slow-relaxing signal
still remains. This gives rise to a gradual decrease
of $V_{\rm ord}$, which reaches zero only at $T \simeq 5$\,K~\cite{Svanidze011026},
in agreement with the results of the wTF measurements. To extract
$T_c$ and the critical exponent $\beta$ of the order parameter,
the $T$-dependence of $B_{\rm int}$ was fitted by the phenomenological
power law $B_{\rm int}(T) = B_{0} \left(1-\frac {T}{T_c}\right)^{\beta}$, 
giving $B_{0} \simeq 14.2$\,mT, $\beta\simeq0.66$, $T_c\simeq2.5$~K.
The obtained $\beta$ is even larger than $0.5$, the
theoretical value expected for a mean-field model. However,
our accuracy of $\beta$-value determination is limited by the low
density of data points near $T_c$. Further, the $T$-dependence
of the obtained transverse $\lambda_{\rm T}$ and longitudinal $\lambda_{\rm L}$
relaxation rates is shown in Fig.~\ref{FigS6}(d). As expected for
a long-range magnetic order, $\lambda_{\rm L}$ shows a peak at $T_c$, whereas, $\lambda_{\rm T}$ shows a continuous growth below it.

\bibliography{NiMo}